# Width of Theta - pentaquarks in relativistic quark model

Gerasyuta S.M.[1,2], Kochkin V.I.[1]

1. Department of Theoretical Physics, St. Petersburg State University, 198904, St. Petersburg, Russia.
2. Department of Physics, LTA, 194021, St. Petersburg, Russia,
   E-mail: gerasyuta@sg6488.spb.edu

## Abstract

The relativistic generalization of Faddeev-Yakubovsky approach is constructed in the form of the dispersion relations. The five-quark amplitudes for the low-lying pentaquarks contain u,d,s-quarks. The poles of these amplitudes determine the masses and widths of Theta-pentaquarks.

I. Introduction

Recent observation of an exotic baryon state with positive strangeness, $\theta^+$ (1540), by LEPS collaboration in Spring – 8[1] and subsequent experiments [2, 3, 4, 5, 6, 7, 8, 9, 10] has raised great interest in hadron physics. This state cannot be an ordinary three-quark baryon since it has positive strangeness, and therefore the minimal quark content is $uudd\bar{s}$. The most striking feature of $\theta^+$ (1540) is that the width is unusually small: $\Gamma < 25$ MeV despite the fact that it lies about 100 MeV above the NK threshold.

The discovery of $\theta^+$ has triggered intensive theoretical studies to understand the structure of the $\theta^+$ [11, 12, 13, 14, 15, 16, 17, 18, 19, 20]. One of the main issues is to clarify the quantum numbers, especially, the spin and the parity, which are key properties to understand the abnormally small width. Recent analysis of the $K^+$ scattering from the xenon or deutron implies even smaller value $\Gamma < 1$ MeV [21, 22, 23]. The chiral soliton model has predicted the masses and widths of the pentaquark baryons with less theoretical ambiguity based on the SU(3) flavor algebra [24]. The model indicates the width of $\theta^+$ around a few ten MeV [25]. In the non-relativistic quark model (A.Hosaka et al) [26] found that the negative parity $\theta^+$ width becomes very large which is of order of several hundreds MeV, while it is about a several tens MeV for the positive parity. By assuming additionally diquark correlations, the width is reduced to be of order 10 MeV.

In the present paper the relativistic generalization of five – quark equations (like Faddeev-Yakubovsky approach) are used in the form of the dispersion relation [27]. The five-quark amplitudes for the low-lying pentaquarks contain u, d, s - quarks. The poles of these amplitudes determine the masses of $\theta^+$ - pentaquarks. The mass spectra of the isotensor $\theta^+$ - pentaquarks with $J^P = \frac{1}{2}^\pm, \frac{3}{2}^\pm$ are calculated. The masses of the constituent u, d, s – quarks coincide with the quark masses of the ordinary baryons in our quark model [28]: $m_{u,d} = 410$ MeV, $m_s = 557$ MeV. The model has only three parameters. The cut-off parameters $\Lambda_{0^+} = 16.5$ and $\Lambda_{1^+} = 20.12$ are similar to paper [29]. The gluon coupling constant $g = 0.456$ is fitted by fixing the pentaquark with the mass M(1540). We obtained the masses and widths of $\theta^+$ pentaquarks with the $J^P = \frac{1}{2}^\pm, \frac{3}{2}^\pm$ and predicted the masses and widths of $\theta^{++}(\theta^0)$ and $\theta^{+++}(\theta^-)$ pentaquarks (Table I).

We use the five-quark amplitudes for the family of exotic $\theta$ - baryons.

Let us represent that some current generates a five-quark system. Their successive pair interactions allow to construct the correct equations for the amplitude. In order to present the amplitude in the form a dispersion relation it is necessary to define the amplitudes of quark-quark interaction. The pair quarks amplitudes $q\bar{q} \to q\bar{q}$ and $qq \to qq$ are calculated in the framework of the dispersion N/D method with the input four-fermion interaction with quantum numbers of the gluon. The regularization of the dispersion integral for the D – function is carried out with the cut-off parameters $\Lambda_n$.

Then one should represent the five-particle amplitude as a sum of ten subamplitudes:

$$A = A_{12} + A_{13} + A_{14} + A_{15} + A_{23} + A_{24} + A_{25} + A_{34} + A_{35} + A_{45} \qquad (1)$$

We need to consider only one subamplitude with the certain pair interaction. We shall construct the five-quark amplitude of four nonstrange quarks and one strange antiquark. The poles of these amplitudes determine the masses of pentaquarks. The solution of the system of five-quark equations are written as:

$$\alpha_i(s) = F_i(s, \lambda_i) / D(s), \qquad (2)$$

where zeros of $D(s)$ determinants define the masses of bound states of pentaquarks. $F_i(s, \lambda_i)$ are the functions of $s$ and $\lambda_i$ (current parameters). The functions $F_i(s, \lambda_i)$ determine the contributions of the subamplitudes to the pentaquark baryon amplitude and allow to obtain the spectroscopic (overlap) factor of the $\theta$ - baryon family. We calculated the spectroscopic factor $f$ (Table II) and used the phase space of reactions: $\theta \to KN$ and $\theta \to K\Delta$ for all states. We consider the empirical formula:

$$\Gamma \sim f^2 \times 200 \text{ MeV}$$

for the pentaquark $\theta^+(1540)$ and calculated width $\Gamma \sim 15$ MeV. Then we calculated the widths of low-lying pentaquarks (Table I) using the masses was obtained in our paper [27]. We use the formula $\Gamma \sim f^2 \times \rho$ [30], there $\rho$ is the phase space (Table II). We take into account the result of the paper [26] width $\Gamma \sim 570$ MeV for the state $\theta^+$ with $J^P = \frac{1}{2}^-$.

The results of calculations allow to consider two $\theta$ pentaquarks as narrow resonances. The width of $\theta^+(1540)$ with $J^P = \frac{1}{2}^+$ and isospin $I_z = 0$ is equal 15 MeV. The width of $\theta^{++}(1575)$ with $J^P = \frac{1}{2}^+$ and isospin $I_z = 1$ is about 30 MeV. We predict the degenerace of $\theta^{+++}$ pentaquarks with $J^P = \frac{1}{2}^+$ and $J^P = \frac{3}{2}^+$. These states have only the weak decays.

We take into account the contribution of $0^+$ and $1^+$ diquarks. In this case the mass of $\theta$ pentaquarks with the positive parity is smaller than the mass of pentaquarks with negative parity and equal spin.

Acknowledgment

The authors would like to thank T.Barnes,D.I.Diakonov,A.Hosaka,Fl.Stancu for useful discussions.

Table I. Low-lying $\theta$ pentaquark masses and widths (MeV)

| Pentaquark | $J^P$ | Mass, MeV | Width, MeV |
|---|---|---|---|
| $\theta^+$ ($udud\bar{s}$) | $\frac{1}{2}^+$ | 1540 | 15 |
| | $\frac{3}{2}^+$ | 1969 | 430 |
| | $\frac{1}{2}^-$ | 1643 | 570 |
| | $\frac{3}{2}^-$ | - | - |
| $\theta^{++}$ ($uuud\bar{s}$) $\theta^0$ ($dddu\bar{s}$) | $\frac{1}{2}^+$ | 1575 | 30 |
| | $\frac{3}{2}^+$ | 1761 | 70 |
| | $\frac{1}{2}^-$ | 1630 | 390 |
| | $\frac{3}{2}^-$ | 1857 | - |
| $\theta^{+++}$ ($uuuu\bar{s}$) $\theta^-$ ($dddd\bar{s}$) | $\frac{1}{2}^+, \frac{3}{2}^+$ | 1727 | - |
| | $\frac{5}{2}^+$ | - | - |
| | $\frac{1}{2}^-, \frac{3}{2}^-$ | 1704 | - |

Parameters of model: quark mass $m_{u,d}$ = 410 MeV, $m_s$ = 557 MeV;
cut-off parameter $\Lambda_{0^+}$ =16.5, $\Lambda_{1^+}$ =20.12; gluon coupling constant $g$ =0.456.

Table II. Spectroscopic factors $f(0^+)$, $f(1^+)$ and phase space $\rho$ of Theta pentaquark family.

| Pentaquark (channels) | $J^P$ | $f(0^+ - diquark)$ | $f(1^+ - diquark)$ | $\rho$ |
|---|---|---|---|---|
| $\theta^+$ ($KN$) | $\frac{1}{2}^+$ | 0.1450 | 0.1330 | 0.0265 $GeV^2$ |
| $\theta^+$ ($KN$) | $\frac{3}{2}^+$ | - | 0.3180 | 0.2910 $GeV^2$ |
| $\theta^+$ ($KN$) | $\frac{1}{2}^-$ | 0.2610 | 0.1170 | 0.4730 |
| $\theta^{++}$ ($KN$) | $\frac{1}{2}^+$ | 0.2550 | 0.1700 | 0.0402 $GeV^2$ |
| $\theta^{++}$ ($KN$) | $\frac{3}{2}^+$ | - | 0.2580 | 0.1390 $GeV^2$ |
| $\theta^{++}$ ($KN$) | $\frac{1}{2}^-$ | 0.2810 | 0.1860 | 0.4610 |
| $\theta^{++}$ ($KN$) | $\frac{3}{2}^-$ | - | 0.2950 | - |

References

1. T.Nakano et al.LEPS Collaboration,Phys.Rev.Lett. 91,012002(2003)
2. V.V.Barmin et al. DIANA Collaboration,Phys.Atom.Nucl.66,1715(2003)
3. S.Stepanyan et al. CLAS Collaboration,Phys.Rev.Lett. 91,252001(2003)
4. V.Kubarovsky et al. CLAS Collaboration, Phys.Rev.Lett. 92,032001(2004)
5. J.Barth et al.SAPHIR Collaboration,Phys.Lett.B572,127(2003)
6. A.E.Asratyan,A.G.Dolgolenko and M.A.Kurbantsev,Phys.Atom.Nucl.67,682(2004)
7. A.Airapetian et al. HERMES Collaboration,Phys.Lett. B585,213(2004)
8. A.Aleev et al.SVD Collaboration hep-ex/0401024
9. S.V.Chekanov.ZEUS Collaboration,hep-ex/0404007
10. H.G.Juengst. CLAS Collaboration,nucl-ex/0312019
11. R.L.Jaffe and F.Wilczek.Phys.Rev.Lett. 91,232003(2003)
12. M.Karliner and H.J.Lipkin.Phys.Lett. B575,249 (2003)
13. S.Sasaki.Phys.Rev.Lett. 93,152001(2004)
14. F.Csicor,Z.Fodor,S.D.Katz and T.G.Kovacs.JHEP,11,070(2003)
15. Shi-Lin Zhu.Phys.Rev.Lett. 91,232002 (2003)
16. C.E.Carlson,C.D.Carone,H.J.Kwee and V.Nazaryan.Phys.Lett. B573,101(2003)
17. S.Capstick,P.R.Page and W.Roberts.Phys.Lett.B570,185(2003)
18. B.Jennings and K.Maltman.hep-ph/0308286
19. A.Hosaka.Phys.Lett.B571,55(2003)
20. S.M.Gerasyuta and V.I.Kochkin.Int.J.Mod.Phys.E12,793(2003)
21. R.N.Cahn and G.H.Trilling.Phys.ReV.D69,011501(2004)
22. A.Sibirtsev,J.Haidenbauer,S.Krewald and U.G.Meissner.hep-ph/0405099
23. R.L.Jaffe and A.Jain,hep-ph/0408046
24. D.Diakonov,V.Petrov and M.V.Polyakov.Z.Phys.A3359,305(1997)
25. R.L.Jaffe.Eur.Phys.J.C35,221(2004)
26. A.Hosaka,M.Oka and  T.Shimozaki,hep-ph/0409102
27 S.M.Gerasyuta and V.I.Kochkin.hep-ph/0310227
28. S.M.Gerasyuta .Z.Phys.C60,689 (1993)
29. S.M.Gerasyuta and V.I.Kochkin.hep-ph/0310225
30. J.J.Dudek and F.E.Close.hep-ph/0311258